\documentclass[prb,showpacs,twocolumn,superscriptaddress]{revtex4}
\usepackage{amsmath,graphicx,latexsym}

\begin{document}

\title{Self-interaction correction with Wannier functions}

\author{Massimiliano Stengel}
\author{Nicola A. Spaldin}

\affiliation{Materials Department, University of California, Santa Barbara,
             CA 93106-5050, USA}

\date{\today}

\begin{abstract}
We describe the behavior of the Perdew-Zunger self-interaction-corrected local
density approximation (SIC-LDA) functional when implemented in a plane-wave 
pseudopotential formalism with Wannier functions.
Prototypical semiconductors and wide-bandgap oxides show a large overcorrection 
of the LDA bandgap.
Application to transition-metal oxides and elements with $d$ electrons is
hindered by a serious breaking of the spherical symmetry, which appears even 
in a closed shell free atom.
Our results indicate that, when all spherical approximations are lifted,
the general applicability of orbital-dependent potentials is very limited
and should be reconsidered in favor of rotationally invariant functionals.
\end{abstract}


\maketitle


\section{INTRODUCTION}

Within the Kohn-Sham approach to density functional theory, the total energy of a 
many-electron system of density $\rho = \rho_{\uparrow} + \rho_{\downarrow}$ is 
generally decomposed into four terms:
\begin{equation}
E^{KS} = T_s + E_H[\rho] + E_{ext} + E_{xc}[\rho_{\uparrow},\rho_{\downarrow}].
\label{kseq}
\end{equation}
These terms describe the kinetic energy of the fictitious set of one-particle 
orbitals $T_s$, the Hartree energy $E_H$, the energy due to the interaction with 
the external potential $E_{ext}$, and the unknown exchange and correlation energy 
$E_{xc}$; the latter is commonly approximated by local or semi-local functionals 
of the density such as the local density (LDA) or generalized gradient (GGA)
approximations. 

Despite the innumerable successes of the LDA and GGA, some serious drawbacks 
exist that prevent the applicability of these methods to a wider range of materials 
and phenomena.
Situations in which these standard functionals lead to qualitatively incorrect physics
include the erroneous prediction of metallicity for magnetic transition metal oxides, 
an inability to localize defect states in solids~\cite{Avezac/Calandra/Mauri:2005} and 
unpaired electrons in water~\cite{Vandevondele/Sprik:2005}, qualitatively incorrect 
metallic transport for single-molecule junctions~\cite{Toher:2005}, inaccurate
redox potentials and charge-transfer reactions~\cite{Kulik/Cococcioni:2006}, 
and unphysical fractionally charged fragments in the molecular dissociation 
limit~\cite{Ruzsinsky/Perdew:2007}.
These failures can be traced, at least in part, to the self-interaction 
error (SIE), which is the spurious interaction of an electron with its 
own Hartree and exchange-correlation potential.
Indeed, in the case of one-electron systems such as the ground state of the
hydrogen atom, $E_H$ and $E_{xc}$ should cancel exactly: 
$$
E_{H}[\rho_{1s}]+E_{xc}[\rho_{1s},0] \quad \text{    should equal    } \quad 0,
$$
but this condition is not fulfilled by approximate exchange-correlation 
functionals such as LDA or GGA.  
While in a many-electron system the notion of self-interaction
is less clear cut, it is commonly accepted that
this same mechanism affects the behavior of strongly localized,
atomic-like orbitals, such as $d$ states in transition metal compounds,
by suppressing or mistreating on-site Coulomb interactions. 
The considerable fundamental and technological interest in $d$-electron 
systems such as high-$T_c$ superconductors and colossal magnetoresistive 
manganites provides a compelling incentive for implementing appropriate
SIE-free density functional methods. 
Interestingly, the SIE has a relatively minor impact on total energies,
but it strongly affects the eigenvalues of the Kohn-Sham Hamiltonian.
In particular, the energy eigenvalue associated with the highest occupied 
orbital usually shows a strong departure from the ionization potential, while 
it should match it exactly within exact DFT~\cite{PhysRevLett.49.1691}.

Attempts to correct the self interaction error can be traced back to 
the seminal paper by Perdew and Zunger (PZ) \cite{Perdew/Zunger:1981}, who 
defined the self-interaction corrected (SIC) exchange-correlation energy,
$E_{xc}^{SIC}$, as 
\begin{equation}
E_{xc}^{SIC} = E_{xc}^{\text{approx}}[\rho_{\uparrow},\rho_{\downarrow}] - 
\sum_{\alpha \sigma} \left(
E_H[\rho_{\alpha\sigma}] + E_{xc}^{\text{approx}}[\rho_{\alpha\sigma},0] \right). 
\label{PZ-SIC-LDA}
\end{equation}
Here $E_{xc}^{\text{approx}}[\rho_{\uparrow},\rho_{\downarrow}]$ is the approximate
(for example LDA or GGA) exchange-correlation energy, and the term within the 
summation is the self-interaction energy of an electron in orbital $\alpha$ 
with spin $\sigma$; $E_H[\rho_{\alpha\sigma}]$ is the self-Coulomb part and 
$E_{xc}^{\text{approx}}[\rho_{\alpha\sigma},0]$ is the self exchange-correlation part.
For isolated atoms, this approach yielded Hamiltonian eigenvalues which
were in surprisingly good agreement with experimental removal energies.

These successes motivated a considerable subsequent effort to
incorporate PZ self-interaction corrections in calculations for solids. 
Unfortunately, however, direct implementation of the PZ functional in
extended systems has proved to be technically non-trivial.
The main issue arises from the fact that the SIC-LDA functional, 
unlike standard Kohn-Sham theories, is not invariant 
with respect to a unitary transformation of the occupied manifold;
in particular, the SIC vanishes for extended Bloch wavefunctions.
Therefore, the first challenge of any implementation is to devise a 
general and physically sound criterion for the choice of 
this unitary transformation, which yields a set of ``local orbitals'' 
(LO), as opposed to the ``canonical orbitals'' (CO) which
are the usual eigenstates of the Hamiltonian with Bloch periodicity.

Soon after the initial work by Perdew and Zunger, Heaton Harrison and Lin
(HHL) recognized that Wannier functions provide an ideal basis for describing the 
localized-delocalized duality of electrons in the full-SIC Hamiltonian
\cite{Heaton/Harrison/Lin:1983}; by implementing SIC-LDA within a LCAO 
basis set HHL found considerable improvement in the solid Ar and
LiCl bandstructures. 
An appealing aspect of the HHL approach is the introduction of a unified
Hamiltonian by means of band projections. This strategy removes the orbital 
dependence of the SIC Hamiltonian and allows for the calculation of
all SIC-LDA eigenvalues for a given $k$ point by one single matrix 
diagonalization.
Furthermore, HHL defined the unitary transformation between the Wannier and Bloch
representation as the one yielding the variational minimum
of the SIC-LDA functional within the usual orbital orthonormality constraints.
Pederson, Heaton and Lin~\cite{Pederson_et_al:1984} later demonstrated that the 
so-called ``localization condition'' is then fulfilled by the localized orbitals 
$\phi_\alpha$ and their associated SIC potentials $\delta V_\alpha$:
\begin{equation}
\langle \phi_\alpha | \delta V_\alpha - \delta V_\beta| \phi_\beta \rangle.
\end{equation}
This means that the Lagrange multiplier matrix enforcing the orthonormality
condition is Hermitian, and it can be indeed diagonalized to
obtain the SIC eigenvalue spectrum, together with a set of 
eigenvectors that correspond to the Bloch-periodic COs.

Svane and Gunnarsson \cite{Svane/Gunnarsson:1990} (SG) and later Szotek,
Temmerman and Winter \cite{Szotek/Temmerman/Winter:1993} (STW) 
applied a fully self-consistent SIC-LDA method to extended systems within 
an LMTO-ASA (linear muffin tin orbital - atomic sphere approximation) 
implementation, obtaining remarkable results for both $d$- and 
$f$-electron materials.
A major pitfall of the SIC functional is that it allows for multiple local
minima, one of these being the trivial solution where all the LOs are 
Bloch-like (i.e. no SIC is applied), and another obvious one being the one 
where all the LOs are Wannier-like;
intermediate (mixed) choices also exist, where some of the LOs are 
Wannier-like, and others keep their itinerant character. 
SG and STW proposed choosing the solution with the lowest total energy
(which corresponds to the absolute minimum of the SIC functional, 
and is consistent with the variational character of the localization 
procedure).
Based on this choice, phase transitions are sometimes observed as a 
function of external parameters (e.g. volume) in which the SIC contribution 
for a given band becomes positive (or negative); 
this crossover between SIC and no SIC is rationalized
in terms of a physically appealing realization of Mott transitions 
(which are driven by the competition between bandwidth and on-site 
Coulomb repulsion). 

Some fundamental problems with the use of a partially Bloch-like
and localized solution have been pointed out, however, by Arai and Fujiwara 
\cite{Arai/Fujiwara:1995} (AF). 
First, the presence of  ``delocalized'' bands to which no SIC is applied 
leads to a severe size-consistency problem when the extended solid is 
considered as the thermodynamic limit of an increasingly large cluster 
where SIC is unavoidably finite~\cite{Arai/Fujiwara:1995}.
Even in regions where the SIC energies are slightly positive the SIC 
potentials remain strongly attractive, so when the cluster volume 
is increased to the thermodynamic limit strong and unphysical changes 
in the eigenvalue spectrum must be expected.
Second, the sign of the SIC energy (and hence whether or not the orbitals are
treated as localized) is sensitive to details such as the parameterization
of the LSDA. 
Since the main aim of the SIC method is to correct for the
self-exchange error, qualitative differences in the electronic
ground state determined solely by minor details of the correlation functional
are, again, physically hard to justify.

Interestingly, AF also discussed the consequences of the sphericalization 
of the SIC potential, which is routinely performed (see SG and STW) within 
LMTO implementations and was also adopted in the early works of
HHL \cite{Heaton/Harrison/Lin:1983}.
While a significant impact on the bandstructure of solids and
unphysical energy splittings within otherwise crystal-symmetric multiplets 
were found, AF concluded that the overall consequences of this 
approximation were relatively unimportant.

It has been shown recently for a wide range of atoms that
orbitals with different angular momenta are allowed to mix
upon lifting the spherical approximation~\cite{Vydrov/Scuseria:2005}. 
For example $3s$ and $3p$ states in Ar mix to yield four tetrahedrally 
symmetrical $sp^3$ hybrids, analogous to the maximally localized Wannier 
functions in $sp$ compounds~\cite{Posternak_et_al:2002}.
As a consequence of this mixing, the SIC energy becomes 
\emph{negative} for all bands, while it is generally positive for pure $p$
states; furthermore, eigenvalue shifts of the order of 1 eV occur, 
and in general the agreement with experiments tends to worsen~\cite{Vydrov/Scuseria:2005}.
These atomic results suggest that, while the sphericalization of the SIC
potential itself has a very minor \emph{direct} impact, in agreement
with the conclusions of Refs.~\onlinecite{Heaton/Harrison/Lin:1983}
and~\onlinecite{Arai/Fujiwara:1995}, it might well have a dramatic 
\emph{indirect} impact, by preventing the true variational minimum of 
the SIC functional from being found.
In particular, if the total energy criterion for the selection of the 
localized/delocalized bands is enforced, the artificial suppression
of interband mixing might lead to erroneous choices, and qualitatively
incorrect electronic ground states; for example, oxygen $2p$ states, that 
are considered to be itinerant within the spherical approximation
\cite{Svane/Gunnarsson:1990,Szotek/Temmerman/Winter:1993}
can become localized once mixing with oxygen $2s$ states is
allowed.

It is apparent from the above analysis that two main issues affect 
self-interaction corrected calculations for extended solids, i) the
existence of multiple local minima and ii) the validity of the spherical
approximation.
In this work we address both issues by using modern 
Wannier-function theory \cite{Marzari/Vanderbilt:1997} and a plane-wave
norm-conserving pseudopotential formalism.
By testing our method on simple atomic systems we first demonstrate that, 
if used with care, the pseudopotential approximation
introduces a negligible error with respect to the most accurate
all-electron results available to date \cite{Vydrov/Scuseria:2005};
this provides a strong validation of our results and
contributes to putting the full-SIC formalism onto a solid
implementation-independent technical footing.
In particular, in agreement with Ref. \onlinecite{Vydrov/Scuseria:2005},
we find that the spherical approximation has an important impact on
the eigenvalue spectrum of solids, often significantly worsening the
agreement with experimental spectroscopic data.
We further find that, within our spherically-unrestricted SIC-LDA, the 
fully localized solution is always the variational electronic ground state, 
even in bulk Si where the valence electrons are usually 
considered as being itinerant; this result suggests that some caution
must be taken when interpretating the localized/delocalized SIC crossover 
in terms of a Mott transition, since it might be an artifact of the
numerical approximations used.
Finally, in our implementation two further pitfalls of the SIC-LDA method 
emerge, which were so far overlooked in the literature: i) the giant
overcorrection of the electronic band gaps in solids and ii) the unphysical 
breakdown of crystal point symmetry, which is especially serious in 
$d$-electron systems.
We rationalize these effects in terms of, respectively, lack of proper 
treatment of dielectric screening and the rotational non-invariance 
of the method.
Our results provide useful guidelines for further research in
the quest for an improved density functional, and also a benchmark 
against which approximate flavors of SIC 
\cite{Vogel/Kruger/Pollmann:1996,Filippetti/Spaldin:2003,Pemmaraju_et_al:2007}  
can be tested and validated.

The remainder of this work is organized as follows. In Sec.~\ref{method} 
we give a detailed overview of the SIC technique we use in this work. 
In Sec.~\ref{results} we present our results: First we validate our
method by performing some tests on simple atomic systems,
then we apply SIC to simple, prototypical solids (Ar, Si, MgO) and 
finally, we analyze the performance of SIC for $d$-electron systems. 
In Sec.~\ref{discussion} we discuss these results in light of
previously reported studies and analyze their impact for future methodological
development. 
In Sec.~\ref{conclude} we summarize and conclude. The Appendix
presents an analysis of Boys orbitals in $d$-electron spherically symmetric
atoms; this analysis extends the work by Posternak~\cite{Posternak_et_al:2002} 
for $sp$ elements, and shows that orbital-dependent functionals tend to 
unphysically break the symmetry of closed-shell atoms.

\section{Method}
\label{method}

Before presenting our method we briefly summarize some basic notations and
conventions that will be useful in the derivation (for a more extensive 
treatment see 
Ref.~\onlinecite{Stengel/Spaldin_PRB:2006} and references therein).
We assume a Born-von K\'arm\'an supercell of
volume $\Omega_{BvK} = N \Omega$, where $N$ is the total number
of $k$ points arranged on a regular three-dimensional mesh and
$\Omega$ is the volume of the primitive cell used to represent the
periodic crystal.
The generalized Bloch orbitals $\psi_{n\mathbf{k}} (\mathbf{r})$ 
(which are not necessarily eigenstates of the Hamiltonian) can be 
written in terms of the cell-periodic functions 
$u_{n \mathbf{k}} (\mathbf{r})$:
$$
\psi_{n\mathbf{k}} (\mathbf{r}) = e^{i \mathbf{k}. \mathbf{r}}
u_{n \mathbf{k}} (\mathbf{r}) \quad;
$$
the latter can be represented in reciprocal space as follows:
$$
\tilde{u}_{n \mathbf{k}} (\mathbf{G}) =
\frac{1}{\sqrt{\Omega}} \int_{cell} e^{-i\mathbf{G.r}}
 u_{n \mathbf{k}} (\mathbf{r}) d\mathbf{r} \quad.
$$
The reciprocal lattice vectors $\mathbf{b}$ of the Born-von K\'arm\'an 
supercell can be written in terms of $\mathbf{G}$ and the $k$-point mesh:
$$
\mathbf{b} = \mathbf{G} + \mathbf{k} \quad,
$$ 
and the Wannier function associated with the band $n$ and the lattice site 
$\mathbf{R}$ in reciprocal space is:
\begin{equation}
\tilde{w}_{\mathbf{R}n}(\mathbf{b}) = \frac{1}{\sqrt{N}}
                 e^{-i\mathbf{b.R}}  \tilde{u}_{n\mathbf{k}}(\mathbf{G}) \quad.
\label{wfgspace}
\end{equation}
We will set $\mathbf{R}=0$ in the remainder
of the paper, and thus focus on the minimal set of Wannier functions which 
is necessary to describe the solid; also we will introduce a spin
index $\sigma$.
Upon Fourier transformation one obtains the Wannier functions 
$w_{n\sigma}(\mathbf{r})$ 
in real space and their associated charge density distributions
$\rho_{n\sigma}(\mathbf{r}) = |w_{n\sigma}(\mathbf{r})|^2$.
The self-interaction energy of the system, which needs to be added to the
LDA (or GGA) total energy, is then given by Eqn.~\ref{PZ-SIC-LDA}.

In order to minimize the SIC-LDA functional we need to calculate gradients 
of the SIC energy with respect to the wavefunction plane-wave coefficients.
We start by calculating the gradients of the SIC energy with respect to the 
Wannier functions, which can be written as:
$$
\frac{ \delta E^{SIC} }{ \delta w_{n\sigma}^*(\mathbf{r}) }  = 
\hat{V}^{SIC}_{n\sigma}(\mathbf{r})  w_{n\sigma} (\mathbf{r}) \quad.
$$
Here $\hat{V}^{SIC}_{n\sigma}$ is the SIC (Hartree plus exchange and correlation) 
potential generated by the Wannier density $\rho_{n\sigma}(\mathbf{r})$.
The state-dependent potential $\hat{V}^{SIC}_{n\sigma}$ can be recast into a unified
operator by using band projections:
\begin{equation}
\hat{V}^{SIC} = \sum_{n\sigma} \hat{V}^{SIC}_{n\sigma} | w_{n\sigma} \rangle 
\langle w_{n\sigma} | \quad .
\end{equation}
In general, $\hat{V}^{SIC}$ has nonzero Hermitian and anti-Hermitian components:
\begin{eqnarray}
\hat{V}^{SIC-H} & = & \frac{1}{2} (\hat{V}^{SIC} + \hat{V}^{{SIC}\dagger}) \\
\hat{V}^{SIC-A} & = & \frac{1}{2} (\hat{V}^{SIC} - \hat{V}^{{SIC}\dagger})  \quad .
\end{eqnarray}
When applied to the electronic wavefunctions, the anti-Hermitian part 
$\hat{V}^{SIC-A}$ produces a unitary mixing within the occupied manifold
and is the signature of the rotational non-invariance of the SIC-LDA 
functional; no such term exists in standard Kohn-Sham theories.
The Hermitian term, on the other hand, evolves the electronic subsystem in a
direction which is perpendicular to the occupied subspace, and can be treated
as an additional term to be added to the Kohn-Sham Hamiltonian.
For reasons of transparency and in order to have better control over the
minimization process we decided to separate the two tasks into two nested
loops.

In an inner loop, we constrain the update of the wavefunctions to 
a unitary mixing within the occupied manifold: 
$$
u_{n \mathbf{k}}' (\mathbf{r}) = \sum_{m}
u_{m \mathbf{k}} (\mathbf{r}) U^{(\mathbf{k})}_{mn} \quad,
$$
and we seek the set of unitary matrices $U^{(\mathbf{k})}$ that 
yields the minimum value of the SIC energy (the standard Kohn-Sham
energy is invariant with respect to this transformation).
This operation is formally analogous \cite{Stengel/Spaldin_PRB:2006} to the 
calculation of the maximally localized Wannier functions for a set of 
entangled energy bands~\cite{Marzari/Vanderbilt:1997}, except that, instead 
of minimizing the quadratic spread, here we need to enforce the representation 
with the minimum value of the Perdew-Zunger self-interaction. 
In particular, the rotation matrices are obtained by adding an infinitesimal
anti-Hermitian matrix to the identity:
$$
U^{(\mathbf{k})} \sim 1 + dW^{(\mathbf{k})} \quad;
$$
the variation of the SIC-LDA functional with 
respect to this transformation is provided by $\hat{V}^{SIC-A}$:
$$
\Big( \frac{d E^{SIC}}{dW^{(\mathbf{k})}} \Big)_{mn} = 
\langle \psi_{ m \mathbf{k}} | \hat{V}^{SIC-A} | \psi_{ n\mathbf{k}} \rangle \quad.
$$
It is then easy to show that the stationarity of the functional 
(zero gradient) implies:
$$
\langle w_m | V_n - V_m | w_n \rangle = 0 \quad,
$$
which is the usual ``localization condition''~\cite{Svane:1996}.

In the outer loop we add to the LDA Hamiltonian the Hermitian part of the SIC 
operator:
$$
\hat{H}^{SIC-LDA} = \hat{H}^{LDA} + \hat{V}^{SIC-H} \quad, 
$$ 
which is now identical to the full SIC operator since the
anti-Hermitian part vanishes within the subspace spanned by the occupied 
bands. We then take standard electronic steps until the ground state is reached.
At self-consistency, the eigenvalues of this SIC Hamiltonian formally 
agree with the eigenvalues of the Lagrange multiplier matrix that can be
obtained within direct minimization techniques~\cite{Goedecker:1997}.

We note that particular care must be taken in the correct evaluation of the 
self-Hartree energy and potential of the Wannier charges, since the periodic 
boundary conditions induce some spurious long-range interactions between the 
localized charge distributions.
While some authors~\cite{Kiril} have proposed to truncate the $1/r$ Coulomb 
interaction to eliminate the divergence for $\mathbf{G}=0$, in this work
we use the standard approach of introducing a uniform background charge to 
neutralize the system: 
\begin{equation}
E^H_0[\rho] = \frac{1}{2}
\frac{4\pi}{\Omega_{BvK}} \sum_{\mathbf{b}\neq 0} 
\frac{|\rho(\mathbf{b})|^2}{b^2} \quad. 
\end{equation}
For a cubic $BvK$ cell of dimension $L$, the error due to the use of
periodic boundary conditions can be corrected up to the order 
$\mathcal{O}(L^{-5})$ by the following term~\cite{Makov/Payne:1995}:
\begin{equation}
E^H_{corr} = \frac{\alpha}{2L} + \frac{2\pi}{3L^3} 
\int_{cell} d^3 r \rho_n(\mathbf{r-r_0}) r^2, 
\end{equation}
where $\mathbf{r_0}$ is the center of the Wannier function charge.
We use the same form for the Hartree potential, which is the analytic
derivative of the above term with respect to the charge density $\rho_n$;
the relationship $E^H = 1/2 \int V^H \rho d^3r$ is exactly respected.

For both the inner and the outer loops we use a damped-dynamics minimization
algorithm. For the former, we checked the internal consistency of the 
implementation by taking a frictionless run; the mathematical constant of 
motion was conserved within machine precision. The latter procedure is
the standard Car-Parrinello approach with the addition of the Hermitian
SIC operator. The method was implemented in an ``in-house'' electronic 
structure code.
For all our tests we used a cubic $BvK$ supercell, the local density 
approximation and norm-conserving pseudopotentials~\cite{fhi_ref}.
The atomic tests were performed by using a $\Gamma$-point only sampling
of the Brillouin zone and a large supercell; the above algorithm did not
require any modifications, since it was constructed to 
be invariant with respect to Brillouin-zone folding, and hence the 
$\Gamma$-point only calculations are just a special case.
This flexibility provides an appealing link between isolated atoms and
solids, which can be treated on the same footing with the exact same
computational parameters and pseudopotentials.

\section{Results}
\label{results}

\subsection{Test: $sp$ atoms, Be and Ar}

In order to check the reliability of our method we first apply our SIC-LDA
functional to simple isolated atoms, with only $s$ and $p$ valence electrons.  
For consistency with the bulk solids, we perform these tests with the same
plane-wave electronic structure code by placing the individual atoms in a large 
cubic cell of dimension $a_0=10$ \AA.
For Be we obtain the eigenvalue $\epsilon_{2s}$ = -9.2 eV (LDA=-5.6 eV), 
compared to -9.1 eV recently obtained with an all-electron SIC-LDA
formalism~\cite{Vydrov/Scuseria:2005}.
For Ar we find the same level of agreement with the all-electron SIC-LDA 
result~\cite{Vydrov/Scuseria:2005}: $\epsilon_{3p}$ = -16.8 eV in our
calculation (LDA=-10.4 eV), compared to -16.7 eV (all-electron).
It is reassuring to see that the pseudopotential frozen-core approximation,
together with the adoption of periodic-boundary conditions, has negligible 
influence on the accuracy of SIC-LDA in atoms, with an error of about
0.1 eV in a contribution that amounts to 3-6 eV. 
This favorable agreement between two very different electronic structure methods 
stems from the fact that the formalism (global minimum of the SIC-LDA functional with 
orthogonality constraints and no spherical averaging) is the same.

We note that, while the Be example is trivial (there is only one spherically
symmetric $s$ orbital, and no optimization of the ``rotational'' internal
degrees of freedom is necessary), the Ar case has a more interesting solution
in that the localized orbitals, just like the Boys orbitals, are four $sp^3$
hybrids with tetrahedral symmetry.
The mixing of $s$ and $p$ orbitals is only allowed when the spherical approximation
is lifted, and has dramatic consequences on orbital eigenvalues~\cite{Vydrov/Scuseria:2005}. 
Indeed, if we artificially suppress this mixing, we obtain $\epsilon_{3p}$ = -15.6 eV,
a value which is quite close to the original Perdew-Zunger work ($\epsilon_{3p}$ = -15.8 eV)
and to the experimental ionization energy IE=15.8 eV.
Even if the use of the spherical approximation tends to bring atomic eigenvalues in
much better agreement with the experimental spectroscopic data, this procedure is
ill-defined for solids and molecules and therefore cannot be used as an ingredient for
a general electronic structure method; for this reason we caution against its use, as
did the authors of Ref.~\onlinecite{Vydrov/Scuseria:2005}.

We also note that, by suppressing the $s-p$ intermixing, the SIC energy associated with
each local orbital changes radically. The four symmetric $sp^3$ hybrids contribute
$\delta_{sp^3}=-0.45$ eV per electron, while in the restricted solution the $s$ orbital
contributes $\delta_{s}=-0.52$ eV and the $p$ orbital $\delta_{p}=0.15$ eV.
In addition to a much higher total energy, the sphericalized 
solution is characterized by a positive value for $\delta_{p}$. 
Thus, the $p$ states might be incorrectly discarded in the variational optimization 
procedure once the atom is embedded in a periodic lattice \cite{Svane/Gunnarsson:1990}, 
when in fact the solution with hybridized $sp^3$ states would have lower energy. 

\begin{figure}
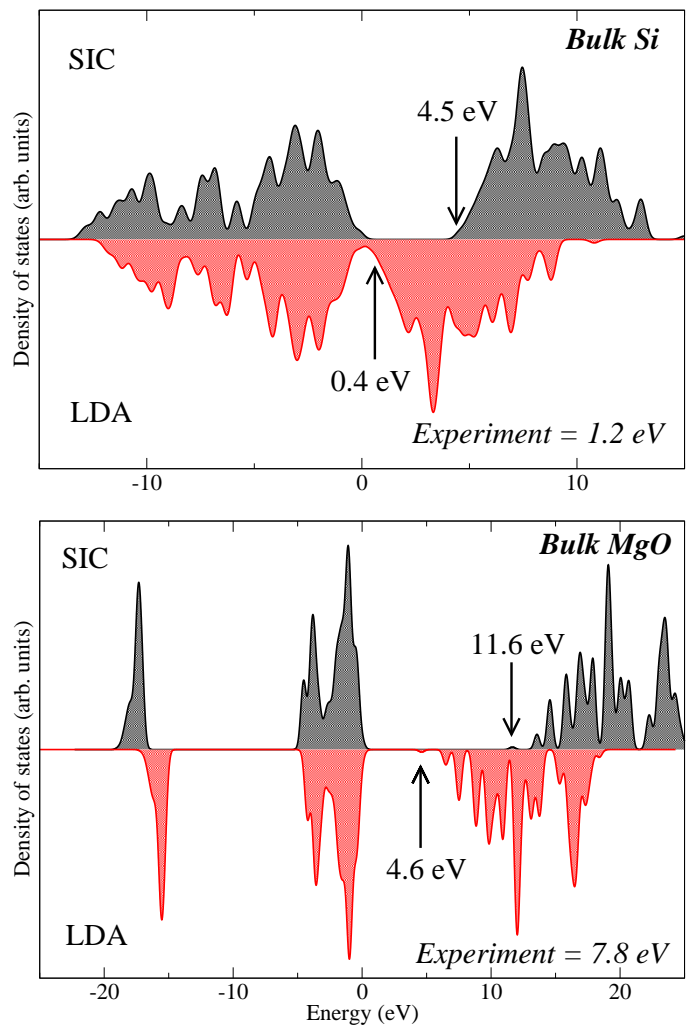

\begin{center}
\includegraphics[width=0.5\textwidth]{si.eps}
\includegraphics[width=0.5\textwidth]{mgo.eps}
\end{center}
\caption{Density of states for bulk Si (top) and MgO (bottom) calculated within
the SIC and LDA approximations. The arrows and numerical values indicate the band 
gaps in the two approximations to be compared with the experimental values at
the bottom right of each plot.}
\label{Si_MgO}
\end{figure}

\subsection{$sp$ Solids: Ar, Si and MgO}

Having assessed the reliability of our plane-wave pseudopotential 
implementation of the Perdew-Zunger SIC functional, we now move on to
the more interesting case of solids. We choose as our examples 
face-centered-cubic (FCC) Ar, to make a direct link to the atomic tests 
reported in the previous subsection, a prototypical semiconductor (Si) and an insulator
(MgO); all these materials show the typical LDA underestimation of the band gap.
The motivation for investigating these apparently ``simple'' compounds is 
to better understand the behavior of the SIC functional in well-known 
test-case systems, before moving to more complex solids where the description 
at the LDA level is highly problematic.
We use a FCC primitive cell and a simple cubic Born-von K\'arm\'an supercell
corresponding to a $5 \times 5 \times 5$ ($3 \times 3 \times 3$) $k$-point meshes 
in the cases of Si (MgO and Ar). 
The experimental lattice constants are used in the Si and MgO cases (10.2 and 7.96 bohr 
respectively), while the lattice constant is progressively varied
in the case of Ar from the experimental value to an artificially compressed
state. Plane-wave cut-offs of 50 Ry, 20 Ry, 70 Ry are used in the respective 
cases of Ar, Si and MgO.

The main goal of the calculations for solid Ar is to quantitatively evaluate
the effect of SIC in the transition from the atomic problem to the bulk solid. 
In particular, we test the statement that ``the SIC-LSD approximation provides a
mechanism which allows the wave functions to localize for systems where
the hopping integrals are small relative to the Coulomb interactions''
\cite{Svane/Gunnarsson:1990}, by tuning the magnitude of the bandwidth, $W$.
%
We use hydrostatic pressure to vary the bandwidth of the $3p$ band of FCC Ar from the 
experimental equilibrium volume ($a_0$ =9.9 a.u., $W_{LDA}$=1.3 eV) to a highly 
compressed state ($a_0$ =7.0 a.u., $W_{LDA}$=8.4 eV), and monitor the effect of SIC
for these extremes together with two intermediate values; the results
are reported in Table~\ref{tab1}.

\begin{table}
\begin{tabular}{|c||c|c||c|c||c|c|c|}
\hline
$a_0$ (bohr)& $W_{LDA}$ & $E_g^{LDA}$ & $W_{SIC}$ &  $E_g^{SIC}$ &
$E_{SIC}$ & $\Delta_{g}$ & $\langle \hat{V}_{SIC} \rangle$ \\
\hline
9.9  & 1.33 &  8.03 & 1.38 & 14.68 & -0.45 & 6.64 & -7.05 \\
9.0  & 2.31 &  8.54 & 2.47 & 15.05 & -0.45 & 6.51 & -7.05 \\
8.0  & 4.38 &  9.75 & 4.67 & 16.25 & -0.45 & 6.50 & -7.07 \\
7.0  & 8.31 & 12.69 & 8.84 & 19.01 & -0.44 & 6.32 & -7.12 \\
\hline
Atom & -    & 10.39 &  -  & 16.85  & -0.46 & 6.46 & -7.07 \\
\hline
\end{tabular}
\caption{Bandwidths, $W$, and band gaps, $E_g$, calculated within the
LDA and SIC approximations for solid Ar over a range of lattice constants.
All energies are reported in eV. The last three columns show the SIC
energy contribution per electron, $E_{SIC}$, the SIC correction to the
band gap $\Delta_g$ and the average of the SIC potential, $\langle V_{SIC} \rangle$.
The lowest row lists the corresponding values for the isolated atom.}
\label{tab1}
\end{table}

As expected, the LDA $3p$ bandwidth ($W_{LDA}$) progressively increases as the crystal is
compressed, and the electronic gap increases as well. The same trend is 
respected in SIC-LDA, with a slightly larger bandwidth (by 0.1-0.5 eV) and 
a rather dramatic opening of the electronic gap with respect to the 
corresponding LDA results. 
The striking fact which is apparent from the Table is that the SIC correction
to the electronic gap is practically \emph{independent} of pressure, and  
amounts exactly to the correction to the $3p$ orbital eigenvalue of the free 
Ar atom (6.46 eV).
Even more striking is the lack of pressure dependence (within numerical error) of both
the SIC energy contribution per electron $E^{SIC}$ (-0.44 to -0.46 eV), 
and the average value of the SIC potential on the corresponding Wannier
function (-7.05 to -7.12 eV); this indicates that in this system the SIC is
substantially insensitive to the bandwidth of the solid, and corrections 
are identical to those calculated in the free atom.
Most notably, $E^{SIC}$ is almost constant. Therefore, application of
the SIC always lowers the variational energy of this system, and there is no
crossover to a hypothetical delocalized solution.
These results (together with the discussion of atomic Ar in the previous
section) strongly suggest that the itinerant character of the oxygen
$2p$ bands reported previously \cite{Svane/Gunnarsson:1990} is a result of 
the spherical approximation adopted therein, rather than an intrinsic physical 
feature of the SIC-LSD method.

Next we move to the cases of Si and MgO. In Fig.~\ref{Si_MgO} we compare the
calculated SIC and LDA densities of states and band gaps for both materials
at the experimental lattice constants. In all cases the top of the valence
band is set to 0 eV. As in the case of solid Ar,
the main effect of the SIC is an important stabilization of the valence bands 
compared to the unoccupied states; otherwise the density of states appears to 
be almost unaffected, apart from a slight increase of the bandwidth within 
SIC compared to the LDA ground state.
The significant opening of the band-gap leads to a dramatic overcorrection
of the LDA value, especially in the case of bulk Si. The band-gaps within
SIC-LDA are respectively 4.5 eV for Si and 11.6 eV for MgO, compared to the
LDA (experimental) values of 0.4 (1.2) eV and 4.6 (7.8) eV.
This behavior might seem surprising at first sight, especially in silicon where
the highly dispersive character of the valence bands leans heavily towards
a delocalized (Bloch) description of the electrons rather than a localized
one.
Our results, however, suggest that even in Si the Wannier functions (which 
in this case are centered along the Si-Si covalent bonds) are localized enough 
to carry a significant SIC ($E_{SIC}=-0.24$ eV, $\langle \hat{V}_{SIC} \rangle=-4.58$ eV); 
this fact further undermines the validity of the SIC-LDA as 
a theory to discriminate between band insulators and Mott insulators.
As a further proof of the localized character of oxygen $p$ bands in solids,
we note that our SIC-LDA solution for bulk MgO shows similar behavior to 
that of the Ar crystal, in that four $sp^3$ hybrids are formed, each with 
decidedly negative values of $E_{SIC}=-0.47$ eV and 
$\langle \hat{V}_{SIC} \rangle=-7.80$ eV.

We are aware of three previous SIC calculations for these materials.
HHL~\cite{Heaton/Harrison/Lin:1983} found a correction of 6 eV
for the bandgap of solid Ar, which is fairly close to our result in spite of
use of the atomic orbital and spherical approximations in the earlier work. 
FCC Ar was also investigated by Szotek, Temmerman and Winter~\cite{Temmerman_Ar}, who 
found an increase of the bandgap of 5.1 eV only, which is in better agreement with our 
results for the spherically restricted atom.
Bulk Si was studied within an approximate bond self-interaction correction by Hatsugai and 
Fujiwara~\cite{Hatsugai/Fujiwara:1988}, who found a very favorable agreement with the
experimental bandstructure, in striking contrast with our results.
These data further highlight the fact that the approximations that have been commonly adopted
in the literature tend to reduce the systematic, sometimes dramatic, overcorrection of the 
LDA bandstructure which is obtained within a rigorous application of the SIC-LDA functional. 

\begin{figure}
\begin{center}
\includegraphics[width=0.4\textwidth]{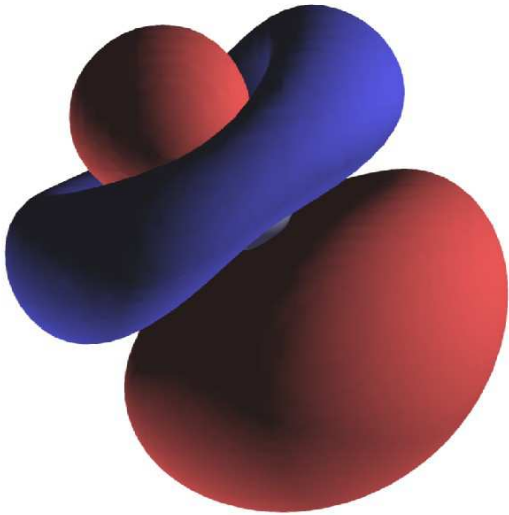}
\end{center}
\caption{A representative $sp^3d^5$ hybrid in the isolated Zn$^{2+}$ ion.}
\label{sp^3d^5}
\end{figure}

\subsection{Materials with $d$ electrons}

\begin{figure}
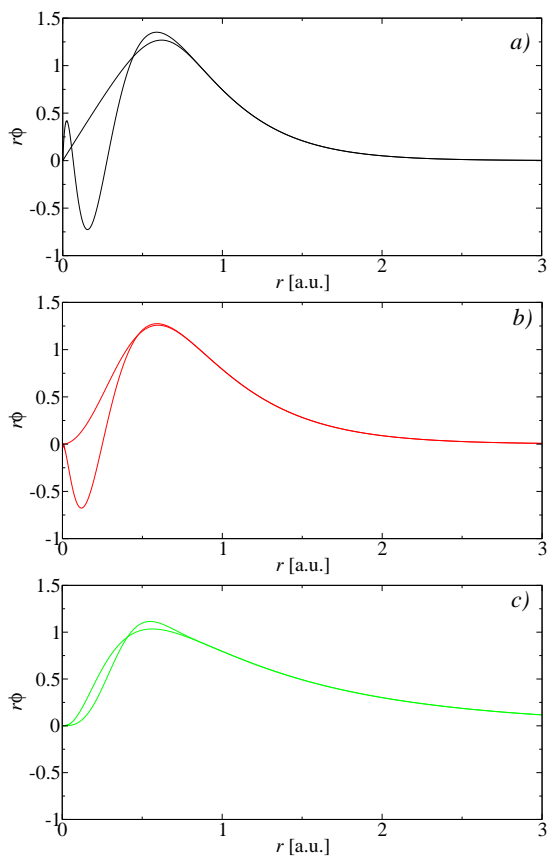

\begin{center}
\includegraphics[width=0.4\textwidth]{ae_ps_0.eps}
\includegraphics[width=0.4\textwidth]{ae_ps_1.eps}
\includegraphics[width=0.4\textwidth]{ae_ps_2.eps}
\end{center}
\caption{All-electron and pseudo orbitals for the isolated, neutral Zn atom. a) $3s$;
b) $3p$; c) $3d$. Note the large spatial overlap between $s$, $p$ and $d$ 
orbitals, whose maximum is located at approximately the same radial distance
from the nucleus.}
\label{PSoverlap}
\end{figure}

As a first step towards studying the effect of SIC on transition metal compounds 
we begin with the case of an isolated $d$-electron atom in a cubic supercell;
this allows us to determine the effect of SIC on $d$ states while
avoiding complications arising from bandwidth and ligands. In particular, we choose for 
simplicity the Zn$^{2+}$ ion, which has a completely filled valence shell. 
Since the Wannier transformation tends to mix wavefunctions that overlap in space, 
it is necessary to include the semicore $3s$ and $3p$ states explicitly as valence 
orbitals; the Zn $3s$, $3p$ and $3d$ orbitals have important spatial overlap (see 
Fig.~\ref{PSoverlap}), in spite of being far from each other in energy. 
As a consequence, the pseudopotentials (Troullier and Martins~\cite{Troullier-Martins}, 
with cutoff radius $r_C$= 1 a.u. for all channels)
are fairly hard and impose a relatively stiff plane wave cutoff of 180 Ry.
We use a cell dimension $a_0=16$ a.u. which is large enough so that the spurious
crystal-field splitting between $e_g$ and $t_{2g}$ orbitals is 
small ($5 \times 10^{-5}$ hartree).

\begin{table}
\begin{tabular}{c|c|c}
\hline
$r$ (bohr)&  $E_{SIC}$ (eV) & $\langle \hat{V}_{SIC} \rangle$ (eV) \\
\hline
\hline
0.5479  & -1.590 &  -17.972 \\
0.5480  & -1.591 &  -17.976 \\
0.6289  & -1.172 &  -16.548 \\
\hline
\end{tabular}
\caption{Radius from the nucleus $r$, SIC
energy contribution per electron, $E_{SIC}$, and the average of the SIC potential, 
$\langle V_{SIC} \rangle$ calculated for the three groups of $sp^3d^5$ hybrids in the isolated Zn$^{2+}$
atom.}
\label{tab2}
\end{table}

\begin{figure}
\begin{center}
\includegraphics[width=0.4\textwidth]{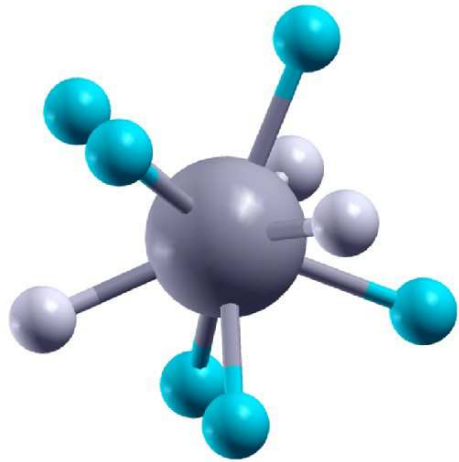} 
\end{center}
\caption{Position of the centers of the $sp^3d^5$ orbitals.}
\label{centersp^3d^5}
\end{figure}

Interestingly, the Wannier localization process of Section~\ref{method}
yields a set of nine similar-looking $sp^3d^5$ hybrids (see Fig.~\ref{sp^3d^5} for
a representative orbital). 
Upon closer inspection of their Wannier centers and their self-interaction energies, 
however, these orbitals differ. In fact they are divided into three groups of three 
members, whose main characteristics are summarized in Tab.~\ref{tab2}.
It is apparent from the Table that the first two groups are practically identical
and in fact they form a group of six which is artificially broken into two by the 
tiny crystal-field splitting imposed by the cubic symmetry of the periodic lattice.
The third group, however, is physically distinct. To visualize the splitting we 
indicate the centers (i.e. the mean value of the position operator) of the orbitals in 
Fig.~\ref{centersp^3d^5}, by highlighting
the members of the group of six as light blue (darker) small spheres 
and the group of three as white (lighter) small spheres; the position of the Zn ion 
is indicated by a larger sphere.

\begin{figure}
\begin{center}
\includegraphics[width=0.4\textwidth]{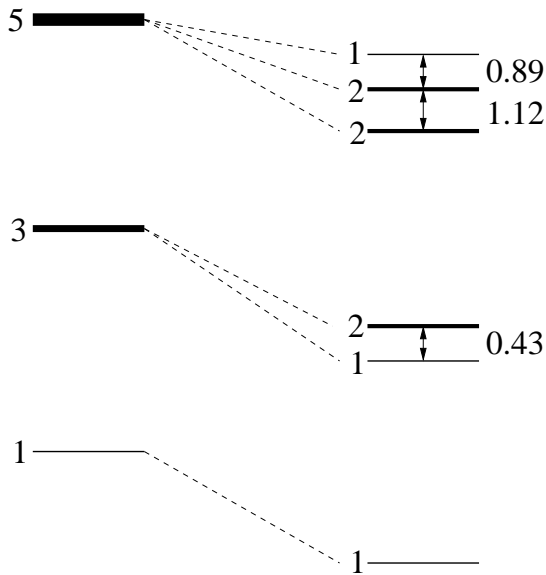} 
\end{center}
\caption{Schematic diagram of the splittings in the eigenvalue spectrum 
of the SIC-LDA Hamiltonian for the Zn$^{2+}$ ion. Values are in eV, the 
spacings are not to scale.}
\label{splittings}
\end{figure}

Unlike the $sp^3$ hybrids, which were characterized by tetrahedral
symmetry and hence did not cause any splitting in the eigenvalue spectrum within the
$3p$ manifold in Ar, the $sp^3d^5$ hybrids of Zn are therefore inequivalent. 
This asymmetry is reflected in the eigenvalue spectrum of the SIC-LDA ground
state of the Zn$^{2+}$ ion, which is represented in Fig.~\ref{splittings}.
The $3p$ multiplet, which lies about 100 eV below the vacuum level,
is split into a doublet and a singlet by 0.43 eV; however the most dramatic 
effect is found in the $3d$ electrons, which are extremely important for
the complex chemistry of the transition metal compounds.
The $3d$ multiplet is split into three levels (2,2,1) by the SIC, with energy 
separation of 1.12 eV and 0.89 eV, i.e. a total of 2 eV between the lowest
and the highest $d$ state. 
This symmetry breaking and consequent splitting of $d$ levels due to the 
non-spherical SIC potential was already pointed out by Arai and Fujiwara in
Ref.  \onlinecite{Arai/Fujiwara:1995}.
That such a drastically unphysical splitting occurs
in a spherically symmetric $d$ electron atom renders the PZ-SIC formalism unreliable 
for $d$ electron solids, where the interplay of crystal field effects
and bandwidth plays a dominant role in determining the overall physical
properties of the compound.

\section{Discussion}
\label{discussion}

Our Wannier basis SIC implementation points to two problems in the PZ-SIC
formalism. The first, the symmetry breaking and splitting of $d$ levels due
to the non-spherical SIC potential is a problem even for the applicatio of
PZ-SIC to atoms. The second, the overestimation of the SIE and consequently
of splitting in the eigenvalue spectrum and band gaps, becomes more acute in
many electron molecules and solids.
Concerning the overestimation of the gap, we argue that to correct the 
LDA bandstructure, not only the local self-interaction of the Wannier 
charges must be taken into account, but also (and especially) the \emph{screening} 
properties of the extended solid upon electron addition/removal; this physical 
ingredient is completely absent in the PZ-SIC-LDA functional, which
is able to capture the dependency on the environment only through the spatial 
distribution of the Wannier charges. 
This is insufficient for a complete picture: 
We have seen in our examples that the effect of the crystal field
on the Wannier densities causes remarkably insignificant 
variations of the SIC correction to the eigenvalue.
In particular, for ionic (or rare gas) solids the individual constituents are 
corrected identically to the isolated ion; this produces a systematic, gross 
overestimation of band gaps.

The overestimation of the band gap points to two interesting and as yet 
unanswered questions regarding 
the physics of the SIE in many-electron systems: What does the self interaction
mean in a many-electron system, and how does it relate to electronic relaxation?
In particular, a theory that is self-interaction free in the Koopman theorem sense,
that is without relaxation corrections, will have over-estimated Hartree-Fock
band gaps in the solid. In fact \emph{the self-interaction 
error is environment-dependent}, and so the relaxation is not 
distinct from SIE but is intimately related to it. 
Rigorous theories to incorporate the dependence of the SIE on the dielectric 
screening environment, such as the GW method, tend to be costly, even if their 
range of applicability is steadily growing~\cite{Neaton/Hybertsen/Louie:2006}. 
Whenever the problem is an ion embedded in a solid with small dispersion
and distinct atomic character, atomic approximations can be quite
effective.
For example, LDA+U has been used recently as an effective technique to cure the SIE,
when the value of the Hubbard U parameter is obtained self-consistently
within a linear-response approach~\cite{Cococcioni/Gironcoli:2005} 
(i.e. it has built-in the dielectric
response of the medium); LDA+U is itself close in spirit to the SIC
approach, although it was derived from a substantially different starting point.
%
%
However, when the covalent character of a given compound is stronger 
(most transition-metal oxides), the reliability of an atomic correction 
applied only to selected bands becomes questionable, hence the interest 
for a more uniform treatment of the occupied bands.

A workaround to this problem within SIC-LDA could be to scale down the SIC 
contribution by a suitable prefactor. This was the approach adopted in the pseudo-SIC
formalism of Filippetti and Spaldin \cite{Filippetti/Spaldin:2003} where
a reduction of the atomic SIC by a factor of 0.5 was included to account
for relaxation effects; within full-SIC, Bylaska, Tsemekhman and Gao~\cite{Kiril} found
that a factor of 0.2 was appropriate to describe defects in Si and C compounds. 
%
%
Recent work for molecules \cite{Vydrov_et_al:2006} showed that a scaling
factor of $\left( \frac{\tau_{\sigma}^{W}}{\tau_{\sigma}} \right)^k$, where
$\tau_{\sigma}$ is the noninteracting kinetic energy density of $\sigma$
spin electrons, and 
$\tau_{\sigma}^{W} = \frac{|\nabla \rho_{\sigma}(r)|^2} {8 \rho_{\sigma}(r)}$
is the Weiz\"{a}cker kinetic energy density, gives improved behavior for atoms and 
molecules.
All of these methods improve the agreement with the experimental bandgaps, 
while still retaining the main physical advantage of SIC: the Hartree 
Fock-like treatment of the on-site Coulomb interactions.

Scaling down the SIC, however, does not remove the unphysical symmetry breaking,
which is especially serious in $d$-electron (and presumably $f$-electron)
materials, and is due to the lack of unitary invariance of the functional.
Therefore we propose that the most promising route to incorporating the
SIC while preserving unitary invariance seems to be the use of hybrid functionals, 
which incorporate a fraction of HF exchange. 
Hybrid functionals have yielded very encouraging results for a wide class 
of systems for the bandgap, structural and electronic properties;
the Wannier function methods presented in this work might be useful in
devising efficient implementations of the Fock exchange within a 
plane-wave pseudopotential formalism.

\section{Conclusions}
\label{conclude}

In summary, we have demonstrated that some common problems of SIC-LDA,
including multiple local minima and size-consistency issues, can be avoided 
by lifting the spherical approximation. However, our calculations expose
two other serious drawbacks of SIC-LDA.
First, we find that the application of the SIC leads to a dramatic overcorrection 
of the electronic bandgap compared to the LDA solution. 
Second, we point out a worrisome, unphysical symmetry 
breaking of spherically symmetric atoms containing $d$ electrons; based on
a perturbative analysis we argue that this drawback might be a general 
feature of state-dependent functionals. 
Our results highlight the deficiencies of state-dependent corrections to 
approximate Kohn-Sham theories, and suggest that rotationally invariant 
corrections (such as the Hartree-Fock exchange in hybrid functionals) are
more promising.

%
%
%
\appendix

\section{Symmetry breaking and Boys localization}

To better understand whether the origin of this breakdown of spherical 
symmetry is related to the special features of SIC-LDA or is a more general
effect that might concern \emph{any} orbital-dependent functional,
it is useful to remove all unnecessary complications and look at the
effect of the simplest possible orbital-dependent perturbation on
the multiplet structure of a spherically symmetric atom with a filled $d$
shell.
A very practical choice is the quadratic spread by Marzari and Vanderbilt, 
which is better known as Boys quadratic spread in isolated atoms and molecules.
Working in the framework of perturbation theory, we start by adding a 
small contribution to the KS total energy:
$$
E^\lambda = E_{KS} + \lambda \Omega, \qquad 
\Omega = \sum_i [\langle r^2 \rangle_i - \mathbf{\bar{r}}^2_i] \quad .
$$
First we need to minimize $\Omega$ with respect to unitary
transformations of the occupied orbitals. We start from a representation 
of angular momentum eigenstates corresponding to the $n=3$ shell of Zn,
i.e. a total of 9 states identified by $l$ and $m$ quantum numbers.
It is clear from the above equation that the minimum spread is
achieved by maximizing the second terms in the square bracket above
(the first is invariant);
these are the diagonal elements of the 3D position operator. 
Equivalently we seek the transformation that \emph{minimizes} the
off-diagonal elements of the three projected position operators, which
do not commute. 
The real-space position operators are particularly simple to calculate on this 
basis, e.g. for $\hat{X}$ we have:
$$
\langle lm |\hat X | l'm' \rangle = \int \phi_{lm}^* (\mathbf{r}) x 
\phi_{l'm'} (\mathbf{r}) d^3r \quad;
$$
$x$ can be written as a real solid harmonic function with $l=1$:
$$
x = -\sqrt{\frac{4\pi}{3}} r Y_{1x}(\mathbf{\hat{r}}) \quad,
$$
so that the above matrix element is simply evaluated in terms of a radial 
integral and a Gaunt coefficient, $G$:
$$
\langle lm |\hat X | l'm' \rangle = -\sqrt{\frac{4\pi}{3}} \int_0^\infty 
r^3 \phi_{lm}^* (r) \phi_{l'm'}(r) dr G^{1x}_{lm,l'm'}.
$$
Angular momentum selection rules yield a set of three sparse matrices with
zero diagonal elements (angular momentum eigenstates are all centered around the
origin) that depend on two values only, which are the $sp$ and 
$pd$ radial overlaps; the solution to the problem, apart from a trivial scaling
factor, is therefore determined by a single parameter.
We used the values from the LDA solution of the isolated Zn pseudoatom
to construct the three matrices. We then
induced a small random unitary mixing to break the symmetry, and we further
optimized the quadratic spread to the minimum until we obtained a set of nine
$sp^3d^5$ localized hybrids. 

By looking at the positions of the centers (radius from the origin) and 
at the individual values of the quadratic spread of each hybrid,
it is clear that the simplified MV spread functional
qualitatively reproduces the same localization pattern 
which was induced by SIC-LDA. 
In particular, the orbitals are split into two inequivalent groups, one of six 
and one of three members; the much smaller splitting of the group of six into two 
subgroups of three (which was observed in our SIC-LDA atomic calculations) is not 
reproduced here since in this case we do not use the supercell approach. 
To appreciate the impact of this perturbation on the Hamiltonian,
we now take the functional derivative of $\Omega$ with respect to 
the wavefunctions; since we are at the variational minimum with respect to
unitary rotations, this functional derivative yields a well-defined
Hermitian operator:
$$
\hat{H}^\lambda = \hat{H}_{KS} + 
\lambda [\hat{R}^2 - 2\sum_i \mathbf{\bar{r}}_i . \langle w_i|\mathbf{\hat{R}}]
\quad .
$$
The $\hat{R}^2$ operator is not orbital dependent, but rather a harmonic 3D
potential acting on all orbitals which preserves spherical symmetry.
The second term in the square bracket, however, does break the symmetry;
in particular, it introduces couplings between states belonging to different
angular momentum multiplets. 
The eigenvalues of this operator reproduce qualitatively the splittings
observed in the SIC-LDA solution, with the 3 degenerate $p$ states
split into 2-1, and the 5 degenerate $d$ states split into 2-2-1.
Therefore, even in the simple case of a harmonic state-dependent perturbation,
the spherical symmetry of the isolated atom is broken, and 
the angular momentum is no longer a good quantum number. This is a very 
undesirable (and artificial) effect that poses serious problems for the 
practical applicability of state-dependent potentials to the physics of 
transition metal oxides.


\bibliography{Nicola}

\end{document}